\documentclass[12pt]{article}
\usepackage{epsfig}
\usepackage{color}
\usepackage{a4}
\usepackage{latexsym}
\usepackage{cite}
\usepackage{bm}

\usepackage{pslatex}
\usepackage[latin1]{inputenc}
\usepackage[T1]{fontenc}

\usepackage{color,colordvi}
\def\colour4colour#1{\Blue{#1}}
\newcommand{\colourcolour}[1]{{\color{blue}{#1}}}

\newcommand{\gsim}{\raisebox{-0.07cm}{$\:\:\stackrel{>}{{\scriptstyle
 \sim}}\:\, $} }

\newcommand{\equal}{\:\: = \:\:}

\newcommand{\hspn}{{\hspace{-4mm}}}
\newcommand{\hspp}{{\hspace{4mm}}}
\newcommand{\beq}{\begin{equation}}
\newcommand{\eeq}{\end{equation}}
\newcommand{\bea}{\begin{eqnarray}}
\newcommand{\eea}{\end{eqnarray}}
\newcommand{\nn}{\nonumber}
\newcommand{\MSb}{$\overline{\mbox{MS}}$}
\newcommand{\ra}{\rightarrow}

\newcommand{\als}{\alpha_{\rm s}}
\newcommand{\ars}{a_{\rm s}}

\newcommand{\ep}{\varepsilon}

\begin{document}
\setlength{\parskip}{0.2cm}
\setlength{\baselineskip}{0.53cm}

\def\nc{{n_c}}
\def\ncs{{n_{c}^{\,2}}}
\def\nct{{n_{c}^{\,3}}}
\def\ncf{{n_{c}^{\,4}}}

\def\mus{{\mu^{\,2}}}

\def\z#1{{\zeta_{#1}}}
\def\zz(#1,#2){{\zeta_{#1}^{\,#2}}}
\def\zss{\zeta_2^{\,2}}

\def\ca{{C^{}_A}}
\def\cas{{C^{\,2}_A}}
\def\cat{{C^{\,3}_A}}
\def\caf{{C^{\,4}_A}}
\def\cf{{C^{}_F}}
\def\cfs{{C^{\, 2}_F}}
\def\cft{{C^{\, 3}_F}}
\def\cff{{C^{\, 4}_F}}
\def\nfz{{n^{\,0}_{\! f}}}
\def\nfo{{n^{\,1}_{\! f}}}
\def\nf{{n^{}_{\! f}}}
\def\nfs{{n^{\,2}_{\! f}}}
\def\nft{{n^{\,3}_{\! f}}}

\def\dfRAnr{{ {d_{\,R}^{\,abcd}\,d_{\,A}^{\,abcd} \over n_c} }}
\def\dfRRnr{{ {d_{\,R}^{\,abcd}\,d_{\,R}^{\,abcd} \over n_c} }}
\def\dabctnr{{ {d^{abc}d_{abc}}\over{n_c} }}

\def\dfAAna{{ {d_{\,A}^{\,abcd}\,d_{\,A}^{\,abcd} \over n_a} }}
\def\dfRAna{{ {d_{\,R}^{\,abcd}\,d_{\,A}^{\,abcd} \over n_a} }}
\def\dfRRna{{ {d_{\,R}^{\,abcd}\,d_{\,R}^{\,abcd} \over n_a} }}

\def\dfRAnA{{ {d_{\,R}^{\,abcd}\,d_{\,A}^{\,abcd} / n_a} }}
\def\dfRRnA{{ {d_{\,R}^{\,abcd}\,d_{\,R}^{\,abcd} / n_a} }}

\def\dfAA{d_A^{\,abcd}d_A^{\,abcd}}
\def\dfFA{d_F^{\,abcd}d_A^{\,abcd}}
\def\dfFF{d_F^{\,abcd}d_F^{\,abcd}}

\def\als{{\alpha_{\rm s}}}
\def\as(#1){{\alpha_{\rm s}^{\:#1}}}
\def\ar(#1){{a_{\rm s}^{\:#1}}}
\def\ars{{a_{\rm s}}}

\def\frct#1#2{\mbox{\large{$\frac{#1}{#2}$}}}

\def\xm{(1\!-\!x)}
\def\LntO{\ln(1\!-\!x)}
\def\Lnt(#1){\ln^{\,#1}(1\!-\!x)}


\begin{titlepage}
\noindent
%
%
DESY-23-150 \hfill October 2023\\    
Nikhef 23-016 \\
LTH 1354 \\
\vspace{0.2cm}
\begin{center}
{\LARGE \bf Additional moments and $\bm{x}$-space approximations\\[1ex]
of four-loop splitting functions in QCD}\\
\vspace{1.3cm}
\large
S. Moch$^{\, a}$, B. Ruijl$^{\,b\,\ast}$, T. Ueda$^{\, c}$, 
J. Vermaseren$^{\, d}$ and A. Vogt$^{\, e}$\\
\vspace{1.2cm}
\normalsize
{\it $^a$II.~Institute for Theoretical Physics, Hamburg University\\
Luruper Chaussee 149, D-22761 Hamburg, Germany}\\
\vspace{3mm}
{\it $^{b}$ETH Z\"urich,
R\"amistrasse 101, CH-8092 Z\"urich, Switzerland}\\
\vspace{3mm}
{\it $^c$
Department of Mathematics, Faculty of Medicine, Juntendo University \\
1-1 Hiraga-gakuendai, Inzai, Chiba 270-1695, Japan}\\
\vspace{3mm}
{\it $^d$Nikhef Theory Group \\
Science Park 105, 1098 XG Amsterdam, The Netherlands} \\
\vspace{3mm}
{\it $^e$Department of Mathematical Sciences, University of Liverpool\\
Liverpool L69 3BX, United Kingdom}\\
\vspace{2.0cm}
{\large \bf Abstract}
\vspace{-0.2cm}
\end{center}
We have extended our previous computations of the even-$N$ moments of 
the flavour-singlet four-loop splitting functions to $N\!=\!12$ for the 
pure-singlet quark case and $N\!=\!10\,$ for all other cases. 
These results, obtained using physical quantities in inclusive
deep-inelastic scattering, have been and will be used to validate
conceptionally much more challenging determinations of these splitting
functions from off-shell operator matrix elements (OMEs).
For~the quark-gluon and gluon-gluon splitting functions, which have yet
to be computed to higher $N$ using OMEs, we construct approximations
based on our moments and endpoint constraints, where we present new
large-$x$ results for the gluon-gluon case. 
These approximations facilitate an approximate N$^3$LO evolution of 
parton distributions which are sufficiently accurate outside the 
region of small momentum fractions~$x$.
 
\vspace*{0.4cm}
$^\ast\:${\small Present address: Ruijl Research, Chamerstrasse 117, 
6300 Zug, Switzerland}
\end{titlepage}
%
%
Parton distribution functions (PDFs) are essential ingredients for all
analyses of hard scattering processes with initial-state hadrons.
Complete calculations of such processes at a certain order in 
renormalization-group improved perturbative QCD require PDFs evolved 
with the corresponding splitting functions.
In particular, complete analyses of benchmark processes at the 
next-to-next-to-next-to-leading order (N$^3$LO), which form an 
important part of the accuracy frontier at the Large Hadron Collider, 
require the four-loop splitting functions $P_{\rm ik}^{\,(3)}(x)$.
The determination of these functions requires very involved
calculations, and only partial results have been obtained so far.

An exact result with a rather direct relevance to phenomenological 
analyses is available so far only for the non-singlet combinations of 
quark distributions \cite{Moch:2017uml}, where the splitting functions
$P_{\rm ns}^{\,(3)}(x)$ are completely known in the (numerically 
relevant) limit of a large numbers of colour $\nc$.
On the other hand, exact results for the N$^3$LO splitting functions 
for the evolution equations
\beq
\label{sgEvol}
  \frac{d}{d \ln\mus} \;
  \Big( \begin{array}{c}
          \! q_{\sf s}^{} \!\! \\ \!g\!
        \end{array}
  \Big)
  \: = \: \left(
    \begin{array}{cc} \! P_{\rm qq} & P_{\rm qg} \!\!\! \\
                      \! P_{\rm gq} & P_{\rm gg} \!\!\! \end{array}
          \right)
  \otimes
  \Big( \begin{array}{c}
          \!q_{\rm s}^{}\!\! \\ \!g\!
        \end{array}
  \Big)
\eeq
for the flavour-singlet quark and gluon PDFs of hadrons,
\beq
\label{sgPDFs}
  q_{\sf s}^{}(x,\mus) \; = \; 
  \sum_{i=1}^{n_{\!f}} \left[\,
     q_i^{}(x,\mus) + \,\bar{q}_i^{}(x,\mus) 
  \:\!\right]
\;\; \mbox{and} \;\; g(x,\mus)
\; ,
\eeq
are presently confined to leading and next-to-leading contributions
in the limit of a large number of light flavours $n_{\!f}$
\cite{Davies:2016jie,Gehrmann:2023cqm,Falcioni:2023tzp}.
Hence N$^3$LO solutions of eq.~(\ref{sgEvol}), where $\otimes$ 
represents the Mellin convolution in the momentum variable, are for now
restricted to approximations based on a limited number of even moments
supplemented by information on the high-energy and threshold limits.

The determination of these moments, which are identical to the 
anomalous dimensions of \mbox{spin-$N$} twist-2 operators up to a 
conventional sign,
\beq
\label{gamP}
  \gamma_{\,\rm ik}^{}(N,\als) \; = \;
  - \int_0^1 \!dx\:\, x^{\,N-1}\, P_{\rm ik}^{}(x,\als)
\; ,
\eeq
can be performed via structure functions in deep-inelastic scattering 
(DIS) or via off-shell operator matrix elements. The former approach 
-- which was pioneered at three loops in 
refs.~\cite{Larin:1993vu,Larin:1996wd} and employed in the complete 
determinations of the N$^2$LO splitting functions 
$P_{\rm ik}^{\,(2)}(x)$ in refs.~\cite{Moch:2004pa,Vogt:2004mw} --
is conceptually simpler, but the complexity of the computations 
scales very unfavourably with $N$.
In~the singlet sector, the latter is conceptually very involved, 
see refs.~\cite{Falcioni:2022fdm,Gehrmann:2023ksf} and references therein, 
but computationally offers a range in $N$ that is roughly twice as 
large as that for the route via DIS.

We have extend our previous computations via four-loop DIS 
\cite{Moch:2021qrk}, performed in {\sc Form} 
\cite{Vermaseren:2000nd,Kuipers:2012rf,Ruijl:2017dtg}
using the {\sc Forcer} program~\cite{Ruijl:2017cxj},
to $N\!=\!12$ for $P_{\rm ps}^{\,(3)} \,=\, P_{\rm qq}^{\,(3)} 
- P_{\rm ns}^{\,(3)+}$ and to $N\!=\!10\,$ for all other
$P_{\rm ik}^{\,(3)}$ in eq.~(\ref{sgEvol}). 
Below we present the analytical results at $N=8$ and $N=10$ in the form
\beq
\label{gamexp}
  \gamma_{\,\rm ik}^{}\left(N,\als\right) \; = \; \sum_{n=0}\,
  \ar(n+1) \,\gamma^{\,(n)}_{\,\rm ik}(x)
\quad \mbox{with} \quad \ars \:\equiv\: \als(\mus)/(4\pi)
\eeq
for a general gauge group. For QCD, i.e., the group $SU(\nc\!=\!3)$,
the basic colour factors are $C_F = 4/3$ and $C_A = \nc = 3$, and the 
quartic group invariants read
$d_{A}^{\,abcd}\,d_{A}^{\,abcd}=135$, 
$d_{\!R}^{\,abcd}\,d_{A}^{\,abcd}=15/2$ and 
$d_{\,R}^{\,abcd}\,d_{\,R}^{\,abcd}=5/12$,
see, e.g. refs.~\cite{vanRitbergen:1998pn,Moch:2018wjh}.
As indicated in eq.~(\ref{gamexp}) we identify, without loss of 
information, the renormalization scale $\mu_{\:\!\sf r}$ with the 
mass-factorization scale $\mu$ of eqs.~(\ref{sgEvol}) and 
(\ref{sgPDFs}).

The N$^3$LO contributions to the pure-singlet anomalous dimensions
in eq.~(\ref{gamexp}) at $N=8,\,10$ are
{\small{
\bea
\label{eq:GpsN8}
 \lefteqn{ \hspn \gamma_{\,\rm ps}^{\,(3)}(N\!=\!8) \equal
%
       \colourcolour{\nf\, \*\cft}\,  \* \left[
            {3960340604223955458923 \over 192072198786048000000}
          - {34718701049 \over 18003384000}\*\zeta_3
          + {13529827 \over 4762800}\*\zeta_4
          - {1369 \over 189}\*\zeta_5
          \right]
} \nn \\[0.5mm] &&  \mbox{\hspn}
       + \,\colourcolour{\nf\, \*\ca}\, \*\cfs\,  \* \left[
          - {43838488788848637899 \over 13719442770432000000}
          + {10167760657 \over 18003384000}\*\zeta_3
          - {10211371 \over 952560}\*\zeta_4
          + {1369 \over 1134}\*\zeta_5
          \right]
\nn \\[0.5mm] && \mbox{\hspn}
       + \,\colourcolour{\nf\, \*\cas}\, \*\cf\,  \* \left[
          - {8552512702477166383 \over 2939880593664000000}
          - {97528710971 \over 18003384000}\*\zeta_3
          + {1340251 \over 170100}\*\zeta_4
          + {128 \over 63}\*\zeta_5
          \right]
\nn \\[0.5mm] && \mbox{\hspn}
       + \,\colourcolour{\nf\, \*\dfRRnr}\,  \* \left[
            {1183211180737 \over 7715736000}
          - {18321694 \over 297675}\*\zeta_3
          - {18164 \over 189}\*\zeta_5
          \right]
\nn \\[0.5mm] && \mbox{\hspn}
       + \,\colourcolour{\nfs\, \*\cfs}\, \* \left[
          - {5115927245667479753 \over 2743888554086400000}
         - {15129691 \over 7144200}\*\zeta_3
          + {1369 \over 1134}\*\zeta_4
          \right]
       + \,\colourcolour{\nfs\, \*\ca\, \*\cf}\, \* \left[
            {15301312238130101 \over 7349701484160000}
\right. \nn \\[0.5mm] && \mbox{\hspn} \left.
          + {8397097 \over 7144200}\*\zeta_3
          - {1369 \over 1134}\*\zeta_4
          \right]
       + \,\colourcolour{\nft\, \*\cf}\, \* \left[
          - {162840799744061 \over 816633498240000}
          + {1369 \over 8505}\*\zeta_3
          \right]
%
\; , \\[2mm]
\label{eq:GpsN10}
 \lefteqn{ \hspn \gamma_{\,\rm ps}^{\,(3)}(N\!=\!10) \equal
%
         \colourcolour{\nf\, \* \cft}\, \* \left[
            {19206657411733877390649313 \over 1118944450162341495000000}
          - {45224548192 \over 28017383625}\*\zeta_3
          + {1080128 \over 539055}\*\zeta_4
          - {25088 \over 5445}\*\zeta_5
          \right]
} \nn \\[0.5mm] &&  \mbox{\hspn}
       + \,\colourcolour{\nf\, \*\ca\, \* \cfs}\, \* \left[
          - {1538138456874500390560463 \over 298385186709957732000000}
          - {31074715888 \over 28017383625}\*\zeta_3
          - {97295744 \over 13476375}\*\zeta_4
          + {12544 \over 16335}\*\zeta_5
          \right]
\nn \\[0.5mm] && \mbox{\hspn}
       + \,\colourcolour{\nf\, \*\cas\, \*\cf}\, \* \left[
          - {202179113304531644762417 \over 284176368295197840000000}
          - {192321673117627 \over 109828143810000}\*\zeta_3
          + {23430848 \over 4492125}\*\zeta_4
          - {14912 \over 49005}\*\zeta_5
          \right]
\nn \\[0.5mm] && \mbox{\hspn}
       + \,\colourcolour{\nf\, \* \dfRRnr} \,\* \left[
            {1240606813603 \over 9901861200}
          - {182828576543 \over 6303268125}\*\zeta_3
          - {1624576 \over 16335}\*\zeta_5
          \right]
\nn \\[0.5mm] && \mbox{\hspn}
       + \,\colourcolour{\nfs\, \* \cfs}\,\*\left[
          - {367710354086746558213 \over 296017050307497750000}
          + {12544 \over 16335}\*\zeta_4
          - {1243744 \over 898425}\*\zeta_3
          \right]
\nn \\[0.5mm] && \mbox{\hspn}
       + \,\colourcolour{\nfs\,\*\ca\,\*\cf}\, \* \left[
            {314242565140920849001 \over 215285127496362000000}
          - {12544 \over 16335}\*\zeta_4
          + {89550464 \over 121287375}\*\zeta_3
          \right]
\nn \\[0.5mm] && \mbox{\hspn}
       + \,\colourcolour{\nft\, \* \cf}\, \* \left[
          - {2205751150439 \over 15885856515375}
          + {25088 \over 245025}\*\zeta_3
          \right]
%
\; .
\eea
}}
The corresponding results for the off-diagonal splitting functions 
are given by
{\small{
\bea
\label{eq:GqgN8}
 \lefteqn{ \hspn \gamma_{\,\rm qg}^{\,(3)}(N\!=\!8) \equal
%
         \, \colourcolour{\nf \* \cft} \*  \left[ 
        -{990917988466579134913309 \over 5762165963581440000000}
      	+{3183230120837 \over 180033840000} \* \zeta_3 
	+{1481184343 \over 47628000} \* \zeta_4
      	+{398159 \over 5670} \* \zeta_5 \right]
} \nn \\[0.5mm] &&  \mbox{\hspn}
       + \, \colourcolour{\nf \* \cfs \* \ca} \*  \left[ 
      	 {173705322188197694847769 \over 411583283112960000000}
	-{5433407245849 \over 60011280000} \* \zeta_3 
	-{417892403 \over 2381400} \* \zeta_4
	-{171271 \over 1620} \* \zeta_5 \right] 
\nn \\[0.5mm] && \mbox{\hspn}
       + \, \colourcolour{\nf \* \cf \* \cas} \*  \left[ 
        -{2068466449111368729523 \over 4899800989440000000}
	-{14832708232003 \over 180033840000} \* \zeta_3 
	+{35445949 \over 108000} \* \zeta_4
	+{23311 \over 140} \* \zeta_5 \right] 
\nn \\[0.5mm] && \mbox{\hspn}
       + \, \colourcolour{\nf \* \cat} \*  \left[ 
          {336616045154933559893 \over 2099914709760000000}
	 +{697606492357 \over 5143824000} \* \zeta_3 
	 -{26056547 \over 141750} \* \zeta_4
	 -{419459 \over 3780} \* \zeta_5 \right] 
\nn \\[0.5mm] && \mbox{\hspn}
       + \, \colourcolour{\nf \* \dfRAna} \*  \left[ 
         -{273996244909 \over 3086294400}
	 -{137047639 \over 396900} \* \zeta_3 
	 +{30298 \over 63} \* \zeta_5 \right] 
\nn \\[0.5mm] && \mbox{\hspn}
       + \, \colourcolour{\nfs \* \cfs} \*  \left[ 
          {40554044566337273617 \over 8231665662259200000}
	 -{26373124409 \over 1285956000} \* \zeta_3 
	 +{127207 \over 11340} \* \zeta_4 \right] 
\nn \\[0.5mm] && \mbox{\hspn}
       + \, \colourcolour{\nfs \* \cf \* \ca} \*  \left[ 
          {36065612407080472327 \over 2939880593664000000}
	 +{5699612263 \over 85730400} \* \zeta_3 
	 -{22867163 \over 680400} \* \zeta_4 \right] 
\nn \\[0.5mm] && \mbox{\hspn}
       + \, \colourcolour{\nfs \* \cas} \*  \left[ 
         -{4518848403845479427 \over 419982941952000000}
	 -{13175860451 \over 285768000} \* \zeta_3 
	 +{15234743 \over 680400} \* \zeta_4
	 +{818 \over 243} \* \zeta_5 \right] 
\nn \\[0.5mm] && \mbox{\hspn}
       + \, \colourcolour{\nfs \* \dfRRna} \*  \left[ 
         -{2023939021 \over 17222625}
	 -{3285578 \over 127575} \* \zeta_3 
	 +{13088 \over 81} \* \zeta_5 \right] 
\nn \\[0.5mm] && \mbox{\hspn}
       + \, \colourcolour{\nft \* \cf} \*  \left[ 
         -{47263236736035329 \over 146994029683200000}
	 +{2244679 \over 1530900} \* \zeta_3 \right] 
       + \, \colourcolour{\nft \* \ca} \*  \left[ 
          {886247558029 \over 1708914965625}
	 -{35816 \over 18225} \* \zeta_3 \right] 
%
\; , \\[2mm]
\label{eq:GqgN10}
 \lefteqn{ \hspn \gamma_{\,\rm qg}^{\,(3)}(N\!=\!10) \equal
%
        \, \colourcolour{\nf \* \cft} \*  \left[ 
          -{774607400252577911077514539 \over 3916305575568195232500000}
	  -{805380500854 \over 140086918125} \* \zeta_3 
	  +{2705671898 \over 94334625} \* \zeta_4
\right.} \nn \\[0.5mm] &&  \mbox{\hspn} \left.
	  +{1934336 \over 16335} \* \zeta_5 \right]
       + \, \colourcolour{\nf \* \cfs \* \ca} \*  \left[ 
           {2957158836400064364217056863 \over 6188729798428752960000000}
	  -{217656420816083 \over 3922433707500} \* \zeta_3 
\right. \nn \\[0.5mm] && \mbox{\hspn} \left.
	  -{15390821408 \over 94334625} \* \zeta_4
	  -{1121272 \over 5445} \* \zeta_5 \right] 
       + \, \colourcolour{\nf \* \cf \* \cas} \*  \left[ 
          -{6335098018460327267287847261 \over 13924642046464694160000000}
\right. \nn \\[0.5mm] && \mbox{\hspn} \left.
	  -{8647744620157 \over 140086918125} \* \zeta_3 
	  +{3177267559 \over 10481625} \* \zeta_4
	  +{312172 \over 1485} \* \zeta_5 \right] 
       + \, \colourcolour{\nf \* \cat} \*  \left[ 
           {683009455461651804853128719 \over 4125819865619168640000000}
\right. \nn \\[0.5mm] && \mbox{\hspn} \left.
	  +{1383109617439853 \over 13312502280000} \* \zeta_3 
	  -{5303419507 \over 31444875} \* \zeta_4
	  -{5029972 \over 49005} \* \zeta_5 \right] 
\nn \\[0.5mm] && \mbox{\hspn}
       + \, \colourcolour{\nf \* \dfRAna} \*  \left[ 
          -{1269333487356283 \over 14522729760000}
	  -{769373679649 \over 2292097500} \* \zeta_3
	  +{7866112 \over 16335} \* \zeta_5 \right] 
\nn \\[0.5mm] && \mbox{\hspn}
       + \, \colourcolour{\nfs \* \cfs} \*  \left[ 
           {4356561239541026269442263 \over 745962966774894330000000}
	  -{287267101372 \over 15565213125} \* \zeta_3 
	  +{269096 \over 27225} \* \zeta_4 \right] 
\nn \\[0.5mm] && \mbox{\hspn}
       + \, \colourcolour{\nfs \* \cf \* \ca} \*  \left[ 
           {698565087254281295546651 \over 73675354743199440000000}
	  +{2781155392789 \over 46695639375} \* \zeta_3 
	  -{133132504 \over 4492125} \* \zeta_4 \right] 
\nn \\[0.5mm] && \mbox{\hspn}
       + \, \colourcolour{\nfs \* \cas} \*  \left[ 
      	  -{261639145927435210838789 \over 24111934279592544000000}
	  -{26791509912217 \over 653738951250} \* \zeta_3
	  +{88731664 \over 4492125} \* \zeta_4
	  +{46688 \over 16335} \* \zeta_5 \right] 
\nn \\[0.5mm] && \mbox{\hspn}
       + \, \colourcolour{\nfs \* \dfRRna} \*  \left[ 
          -{181205970624529 \over 1815341220000}
	  -{2257851248 \over 100051875} \* \zeta_3
	  +{747008 \over 5445} \* \zeta_5 \right] 
\nn \\[0.5mm] && \mbox{\hspn}
       + \, \colourcolour{\nft \* \cf} \*  \left[ 
          -{2121999454705273487 \over 9785687613471000000}
	  +{53744464 \over 40429125} \* \zeta_3 \right] 
       + \, \colourcolour{\nft \* \ca} \*  \left[ 
           {5513232141828253 \over 12708685212300000}
	  -{430256 \over 245025} \* \zeta_3 \right] 
%
\quad
\eea
and
\bea
\label{eq:GgqN8}
 \lefteqn{ \hspn \gamma_{\,\rm gq}^{\,(3)}(N\!=\!8) \:\equal\:
         \colourcolour{\cff} \* \left[ 
       	  - {24543597526270550844943 \over 161340646980280320000} 
          - {15301069838573 \over 63011844000}\*\zeta_3 
	  + {10747 \over 168}\*\zeta_4 
	  + {213596 \over 567}\*\zeta_5 \right]
} \nn \\[0.5mm] &&  \mbox{\hspn}
       + \colourcolour{\cft \* \ca} \* \left[ 
       	    {6310523848625645534759 \over 24009024848256000000} 
          + {15872437061239 \over 63011844000}\*\zeta_3 
	  - {291860591 \over 4762800}\*\zeta_4 
          - {2286556 \over 3969}\*\zeta_5 \right]
\nn \\[0.5mm] && \mbox{\hspn}
       + \colourcolour{\cfs \* \cas}\* \left[ 
       	  - {87368680167927504727 \over 2572395519456000000} 
          + {291378678971 \over 14002632000}\*\zeta_3 
	  - {31798217 \over 1587600}\*\zeta_4 
	  + {4982 \over 21}\*\zeta_5 \right]
\nn \\[0.5mm] && \mbox{\hspn}
       + \colourcolour{\cf\*\cat}\* \left[ 
          - {133262292247883142773 \over 1469940296832000000} 
          - {1106936590687 \over 18003384000}\*\zeta_3 
	  + {5161112 \over 297675}\*\zeta_4 
	  - {73193 \over 7938}\*\zeta_5 \right]
\nn \\[0.5mm] && \mbox{\hspn}
       + \colourcolour{\dfRAnr}\* \left[ 
       	  - {273996244909 \over 4320812160} 
	  - {137047639 \over 555660}\*\zeta_3 
          + {151490 \over 441}\*\zeta_5 \right]
\nn \\[0.5mm] && \mbox{\hspn}
       + \colourcolour{\nf\*\cft}\* \left[ 
       	  - {50125745829108214529 \over 1371944277043200000} 
          - {292991233 \over 150028200}\*\zeta_3 
	  - {580927 \over 39690}\*\zeta_4 
	  + {2960 \over 63}\*\zeta_5 \right]
\nn \\[0.5mm] && \mbox{\hspn}
       + \colourcolour{\nf\*\cfs\*\ca}\* \left[ 
       	    {77269186724872290241 \over 2057916415564800000} 
          - {6495831967 \over 64297800}\*\zeta_3 
	  + {22993211 \over 476280}\*\zeta_4 
	  - {1480 \over 189}\*\zeta_5 \right]
\nn \\[0.5mm] && \mbox{\hspn}
       + \colourcolour{\nf\*\cf\*\cas}\* \left[ 
            {89928989976567629 \over 32665339929600000} 
          + {8867614661 \over 85730400}\*\zeta_3 
	  - {16022087 \over 476280}\*\zeta_4 
	  - {58420 \over 1701}\*\zeta_5 \right]
\nn \\[0.5mm] && \mbox{\hspn}
       + \colourcolour{\nf\*\dfRRnr}\* \left[ 
       	  - {2023939021 \over 24111675} 
	  - {3285578 \over 178605}\*\zeta_3 
          + {65440 \over 567}\*\zeta_5 \right]
       + \colourcolour{\nfs\*\cfs}\* \left[ 
       	    {269507034349861709 \over 102895820778240000} 
\right. \nn \\[0.5mm] && \mbox{\hspn} \left.
	  + {10846 \over 1215}\*\zeta_3
          - {1184 \over 189}\*\zeta_4 \right]
       + \colourcolour{\nfs\*\cf\*\ca}\* \left[ 
       	  - {383192836407971 \over 174992892480000} 
	  - {349492 \over 59535}\*\zeta_3
          + {1184 \over 189}\*\zeta_4 \right]
\nn \\[0.5mm] && \mbox{\hspn}
       + \colourcolour{\nft\*\cf}\* \left[ 
       	    {1517401222367 \over 1701319788000} 
	  - {1184 \over 1701}\*\zeta_3 \right]
 \; , \\[2mm]
\label{eq:GgqN10}
 \lefteqn{ \hspn \gamma_{\,\rm gq}^{\,(3)}(N\!=\!10) \:\equal\:
         \colourcolour{\cff}\* \left[ 
       	  - {10853673038701241183091951323 \over 62660889209091123720000000}
          - {726934765981684 \over 2941825280625}\*\zeta_3 
	  + {16751072 \over 259875}\*\zeta_4
\right.} \nn \\[0.5mm] &&  \mbox{\hspn}\left.
          + {19310144 \over 49005}\*\zeta_5 \right]
       + \colourcolour{\cft\*\ca}\* \left[ 
            {542937177948206382842582969 \over 1670957045575763299200000}
          + {31493864059633 \over 108956491875}\*\zeta_3 
\right. \nn \\[0.5mm] && \mbox{\hspn} \left.
	  - {6864965416 \over 94334625}\*\zeta_4
          - {6174656 \over 9801}\*\zeta_5 \right]
       + \colourcolour{\cfs\*\cas}\* \left[ 
          - {176439961207867804097579923 \over 1670957045575763299200000} 
\right. \nn \\[0.5mm] && \mbox{\hspn} \left.
	  - {579719752264 \over 28017383625}\*\zeta_3
          - {4242844 \over 1164625}\*\zeta_4 
	  + {4582256 \over 16335}\*\zeta_5 \right]
       + \colourcolour{\cf\*\cat} \*  \left[ 
          - {1523068185119005263091080961 \over 27849284092929388320000000}
\right. \nn \\[0.5mm] && \mbox{\hspn} \left.
          - {5408146393855729 \over 109828143810000}\*\zeta_3 
	  + {1127996644 \over 94334625}\*\zeta_4
          - {523568 \over 29403}\*\zeta_5 \right]
       + \colourcolour{\dfRAnr}\* \left[ 
          - {1269333487356283 \over 21784094640000}
\right. \nn \\[0.5mm] && \mbox{\hspn} \left.
          - {769373679649 \over 3438146250}\*\zeta_3 
	  + {15732224 \over 49005}\*\zeta_5 \right]
       + \colourcolour{\nf\*\cft}\* \left[ 
          - {56013386247927639171031907 \over 2237888900324682990000000}
\right. \nn \\[0.5mm] && \mbox{\hspn} \left.
          - {365131853104 \over 46695639375}\*\zeta_3 
	  - {652192 \over 49005}\*\zeta_4 
	  + {3584 \over 99}\*\zeta_5 \right]
       + \colourcolour{\nf\*\cfs\*\ca}\* \left[ 
            {39336775666814707823343377 \over 1491925933549788660000000}
\right. \nn \\[0.5mm] && \mbox{\hspn} \left.
          - {1973089362212 \over 28017383625}\*\zeta_3 
	  + {179052352 \over 4492125}\*\zeta_4 
	  - {1792 \over 297}\*\zeta_5 \right]
       + \colourcolour{\nf\*\cf\*\cas}\* \left[ 
          - {35738891348053198625123 \over 45209876774236020000000}
\right. \nn \\[0.5mm] && \mbox{\hspn} \left.
          + {76945028642074 \over 980608426875}\*\zeta_3 
	  - {357804256 \over 13476375}\*\zeta_4
          - {1291648 \over 49005}\*\zeta_5 \right]
       + \colourcolour{\nf\*\dfRRnr}\* \left[ 
          - {181205970624529 \over 2723011830000}
\right. \nn \\[0.5mm] && \mbox{\hspn} \left.
          - {4515702496 \over 300155625}\*\zeta_3 
	  + {1494016 \over 16335}\*\zeta_5 \right]
       + \colourcolour{\nfs\*\cfs}\* \left[ 
            {7082606576237379049 \over 2990071215227250000}
          + {4834432 \over 735075}\*\zeta_3 
	  - {7168 \over 1485}\*\zeta_4 \right]
\nn \\[0.5mm] && \mbox{\hspn}
       + \colourcolour{\nfs\*\cf\*\ca}\* \left[ 
       	  - {74338063361300467 \over 38126055636900000}
          - {8906512 \over 2205225}\*\zeta_3 
	  + {7168 \over 1485}\*\zeta_4 \right]
\nn \\[0.5mm] && \mbox{\hspn}
       + \colourcolour{\nft\*\cf}\* \left[ 
            {1730630298974 \over 2269408073625} 
	  - {7168 \over 13365}\*\zeta_3 \right]
\; .
\eea
}}
%
%
Finally, the $N=8$ and $N=10$ moments (\ref{gamP}) of the four-loop 
gluon-gluon splitting function read
{\small{
\bea
\label{eq:GggN8}
 \lefteqn{ \hspn \gamma_{\,\rm gg}^{\,(3)}(N\!=\!8) \:\equal\:
         \colourcolour{\caf}\* \left[ 
       	    {913169436152027903 \over 1171827405000000} 
          + {51527099041 \over 131220000}\*\zeta_3 
	  - {172436 \over 405}\*\zeta_5 \right]
} \nn \\[0.5mm] &&  \mbox{\hspn}
       + \colourcolour{\dfAAna}\* \left[ 
       	    {537502950787 \over 551124000} 
	  + {46340503 \over 13500}\*\zeta_3
          - {689744 \over 135}\*\zeta_5 \right]
\nn \\[0.5mm] && \mbox{\hspn}
       + \colourcolour{\nf\*\cft}\* \left[ 
       	  - {3875574534972929574389 \over 576216596358144000000} 
          - {135851486693 \over 18003384000}\*\zeta_3 
	  - {13529827 \over 4762800}\*\zeta_4 
	  + {613 \over 189}\*\zeta_5 \right]
\nn \\[0.5mm] && \mbox{\hspn}
       + \colourcolour{\nf\*\cfs\*\ca}\* \left[ 
       	    {7722227993192582836559 \over 41158328311296000000} 
          + {591286208311 \over 1125211500}\*\zeta_3 
	  + {10006243 \over 952560}\*\zeta_4 
	  - {138355 \over 162}\*\zeta_5 \right]
\nn \\[0.5mm] && \mbox{\hspn}
       + \colourcolour{\nf\*\cf\*\cas}\* \left[ 
          - {793554238929989675839 \over 2939880593664000000} 
          + {534393934321 \over 1800338400}\*\zeta_3 
	  - {40892093 \over 85050}\*\zeta_4 
	  + {320660 \over 567}\*\zeta_5 \right]
\nn \\[0.5mm] && \mbox{\hspn}
       + \colourcolour{\nf\*\cat}\* \left[ 
          - {31340067299523269041 \over 69997156992000000}
          - {680081584643 \over 857304000}\*\zeta_3 
	  + {1788457 \over 3780}\*\zeta_4 
	  + {351572 \over 1701}\*\zeta_5 \right]
\nn \\[0.5mm] && \mbox{\hspn}
       + \colourcolour{\nf\*\dfRAna}\* \left[ 
            {4208909173201 \over 7715736000} 
	  + {3518370631 \over 4465125}\*\zeta_3
          - {1069268 \over 567}\*\zeta_5 \right]
\nn \\[0.5mm] && \mbox{\hspn}
       + \colourcolour{\nfs\*\cfs}\* \left[ 
          - {95246090820533670043 \over 8231665662259200000} 
          + {445763281 \over 21432600}\*\zeta_3 
	  - {6623 \over 5670}\*\zeta_4 \right]
\nn \\[0.5mm] && \mbox{\hspn}
       + \colourcolour{\nfs\*\cf\*\ca}\* \left[ 
            {942217912695786851 \over 14699402968320000}
          - {4533011177 \over 21432600}\*\zeta_3 
	  + {247192 \over 2835}\*\zeta_4 \right]
\nn \\[0.5mm] && \mbox{\hspn}
       + \colourcolour{\nfs\*\cas}\* \left[ 
          - {997206831355739 \over 388873094400000} 
	  + {111733544 \over 535815}\*\zeta_3
          - {162587 \over 1890}\*\zeta_4 
	  - {6280 \over 243}\*\zeta_5 \right]	       
\nn \\[0.5mm] && \mbox{\hspn}
       + \colourcolour{\nfs\*\dfRRna}\* \left[ 
            {331056293 \over 984150} 
	  + {17535248 \over 25515}\*\zeta_3  
          - {100480 \over 81}\*\zeta_5 \right] 
\nn \\[0.5mm] && \mbox{\hspn}
       + \colourcolour{\nft\*\cf}\* \left[ 
          - {12373917501191 \over 30245685120000} 
	  - {1369 \over 8505}\*\zeta_3 \right]
       + \colourcolour{\nft\*\ca}\* \left[ 
          - {263132873693 \over 243045684000} 
	  + {7744 \over 1215}\*\zeta_3 \right]
 \; , \\[2mm]
\label{eq:GggN10}
 \lefteqn{ \hspn \gamma_{\,\rm gg}^{\,(3)}(N\!=\!10) \:\equal\:
         \colourcolour{\caf}\* \left[ 
            {22456657892477562064767080711 \over 25270646676917407920000000}
          + {73846513655083 \over 142369816050}\*\zeta_3 
	  - {153025522 \over 266805}\*\zeta_5 \right]
} \nn \\[0.5mm] &&  \mbox{\hspn}
       + \colourcolour{\dfAAna}\* \left[ 
            {2465616588411197 \over 1815341220000} 
          + {3181991394023 \over 700363125}\*\zeta_3 
	  - {612102088 \over 88935}\*\zeta_5 \right]
\nn \\[0.5mm] && \mbox{\hspn}
       + \colourcolour{\nf\*\cft}\* \left[ 
          - {941134719170744230384913 \over 79924603583024392500000}
          - {4953774597752 \over 980608426875}\*\zeta_3 
	  - {1080128 \over 539055}\*\zeta_4 
	  + {256 \over 135}\*\zeta_5 \right]
\nn \\[0.5mm] && \mbox{\hspn}
       + \colourcolour{\nf\*\cfs\*\ca}\* \left[ 
            {323084460491834273793113701 \over 1491925933549788660000000}
          + {79257043186949 \over 140086918125}\*\zeta_3 
	  + {96076064 \over 13476375}\*\zeta_4 
\right. \nn \\[0.5mm] && \mbox{\hspn} \left.
          - {35324984 \over 38115}\*\zeta_5 \right]
       + \colourcolour{\nf\*\cf\*\cas}\* \left[ 
          - {2224242538139003360626622957 \over 6962321023232347080000000}
          + {419373367171051 \over 1248047088750}\*\zeta_3 
\right. \nn \\[0.5mm] && \mbox{\hspn} \left.
	  - {16266111506 \over 31444875}\*\zeta_4 
          + {69670324 \over 114345}\*\zeta_5 \right]
       + \colourcolour{\nf\*\cat}\* \left[ 
          - {3486469494115338162974489 \over 7032647498214492000000} 
\right. \nn \\[0.5mm] && \mbox{\hspn} \left.
          - {32293176652422641 \over 36609381270000}\*\zeta_3 
	  + {8873246 \over 17325}\*\zeta_4 
          + {81468964 \over 343035}\*\zeta_5 \right]
\nn \\[0.5mm] && \mbox{\hspn}
       + \colourcolour{\nf\*\dfRAna}\* \left[ 
            {4251405683983949 \over 7261364880000}
          + {4659512350937 \over 6303268125}\*\zeta_3 
	  - {214355392 \over 114345}\*\zeta_5 \right]
\nn \\[0.5mm] && \mbox{\hspn}
       + \colourcolour{\nfs\*\cfs}\* \left[ 
          - {20853686824119167585477 \over 1776102301844986500000}
          + {2464070896 \over 121287375}\*\zeta_3 
	  - {61376 \over 81675}\*\zeta_4 \right]
\nn \\[0.5mm] && \mbox{\hspn}
       + \colourcolour{\nfs\*\cf\*\ca}\* \left[ 
            {21257339627304028801747 \over 301399178494906800000}
          - {38450577988 \over 169802325}\*\zeta_3 
	  + {53669108 \over 571725}\*\zeta_4 \right]
\nn \\[0.5mm] && \mbox{\hspn}
       + \colourcolour{\nfs\*\cas}\* \left[ 
            {263981382797132507279 \over 121777445856528000000}
          + {917422934833 \over 3962054250}\*\zeta_3 
	  - {17746492 \over 190575}\*\zeta_4 
	  - {608896 \over 16335}\*\zeta_5 \right]
\nn \\[0.5mm] && \mbox{\hspn}
       + \colourcolour{\nfs\*\dfRRna}\* \left[ 
            {35604721980361 \over 74095560000} 
          + {62420775464 \over 60031125}\*\zeta_3 
	  - {9742336 \over 5445}\*\zeta_5 \right]
\nn \\[0.5mm] && \mbox{\hspn}
       + \colourcolour{\nft\*\cf}\* \left[ 
          - {4047476058908941 \over 8737221083456250} 
	  - {25088 \over 245025}\*\zeta_3 \right]
       + \colourcolour{\nft\*\ca}\* \left[ 
          - {983912572928041 \over 847245680820000} 
	  + {215128 \over 31185}\*\zeta_3 \right]
\; .
\eea
}}
The corresponding result for $N = 2,\,4$ and 6 can be found in 
eqs.~(5)$\,$-$\,$(16) of ref.~\cite{Moch:2021qrk}.

The analytic $N$-dependence of the $\nft$ parts of all four anomalous 
dimensions in eqs.~(\ref{eq:GpsN8})$\,$-$\,$(\ref{eq:GggN10}) has 
been presented in ref.~\cite{Davies:2016jie}, for earlier partial 
results see refs.~\cite{Gracey:1996ad,Bennett:1997ch}. 
Very recently, corresponding results have been obtained for the 
$\nfs$ contributions to $\gamma_{\,\rm ps}^{\,(3)}(N)$ and
$\gamma_{\,\rm gq}^{\,(3)}(N)$ 
\cite{Gehrmann:2023cqm,Falcioni:2023tzp}.
The complete contributions proportional to the value 
$\z4 = \pi^{\,4}/90$ of the Riemann $\zeta$-function have been 
derived for all 4-loop splitting functions in eq.~(\ref{sgEvol}) in 
ref.~\cite{Davies:2017hyl} from the no-$\pi^{\:\!2}$ theorem 
\cite{Jamin:2017mul,Baikov:2018wgs}. 

The coefficients of the quartic group invariants in 
eqs.~(\ref{eq:GpsN8})$\,$-$\,$(\ref{eq:GggN10}) agree with those
that we obtained to higher values in $N$ in ref.~\cite{Moch:2018wjh} 
using off-shell operator matrix elements (OMEs) where we found 
-- using a subset of our DIS-based results \cite{Moch:2021qrk} -- 
relations that allowed us to circumvent the issues concerning their 
renormalization. The present results have been used to check the
OME results in refs.~\cite{Falcioni:2023luc,Falcioni:2023vqq},
where the results for $\gamma_{\,\rm ps}^{\,(3)}$ and
$\gamma_{\,\rm qg}^{\,(3)}$ have been extended to $N = 20$.
Therefore we refrain from writing down our result for 
$\gamma_{\,\rm ps}^{\,(3)}(N\!=\!12)$, it agrees with eq.~(A.8)
of ref.~\cite{Falcioni:2023luc}.

From now on, we focus on the lower-row quantities 
$P_{\rm gq}^{\,(3)}$ and $P_{\rm gg}^{\,(3)}$ which have not been
addressed by OME calculations so far. 
The situation for these is now similar to the 3-loop case before the 
results of ref.~\cite{Moch:2004pa,Vogt:2004mw}, when approximations 
based on a rather small number of moments (then six) were found to be 
adequate over a rather wide range in $x$, cf.~fig.~4 of 
ref.~\cite{vanNeerven:2000wp}. 

In order to make the most of the even moments $2 \leq N \leq 10$, 
we need to supplement these with additional results for $x \ra 1$ and
$x \ra 0$. $P_{\rm gq}^{\,(3)}(x)$ shows a double logarithmic
enhancement in the former limit; the coefficients of 
$\Lnt(\ell \,\geq\, 4)$ have been derived in two different ways in 
refs.~\cite{Soar:2009yh,Almasy:2010wn}. The latter also provides an 
all-order extension of the next-to-next-to-leading logarithmic (NNLL)
contributions which were cast in an analytic form in 
ref.~\cite{Almasy:2015dyv}. 
Disregarding terms that vanish for $x\!\ra\!0$, the large-$x$ behaviour
of the \MSb-scheme diagonal entries in eq.~(\ref{sgEvol}) can be 
written as
\beq
\label{eq:xto1}
  P_{\,\rm kk,\,x\ra 1\,}^{\,(n-1)}(x) \:\: = \:\;
        \frac{A_{\rm k}^{(n)}}{(1-x)_+}
  \,+\, B_{\rm k}^{\,(n)} \, \delta (1\!-\!x)
  \,+\, C_{\rm k}^{\,(n)} \, \LntO
  \,-\, A_{\rm k}^{(n)} + D_{\rm k}^{\,(n)}
\:\: . 
\eeq
Here the coefficients $A_{\,\rm k}^{\,(n)}$ are identical 
\cite{Korchemsky:1988si,Albino:2000cp} to the (lightlike) cusp 
anomalous dimensions which are now fully known at four loops 
\cite{Henn:2019swt,vonManteuffel:2020vjv}.
The coefficients $C^{\,(n)}$ and $D^{\,(n)}$ in eq.~(\ref{eq:xto1}) 
do not contain new information, but are functions of lower-order 
quantities \cite{Dokshitzer:2005bf,Moch:2017uml}. At four loops they 
are given~by
\beq
\label{CD4ofAB}
  C_{\rm k}^{\,(4)} \;=\; \big( A_{\rm k}^{(2)} \big)^{2} 
  \,+\, 2\,A_{\rm k}^{(1)} A_{\rm k}^{(3)}
\;\; , \quad
  D_{\rm k}^{\,(4)} \;=\; \sum_{n=1}^{3} \, A_{\rm k}^{(n)}
  \, \big( B_{\rm k}^{\,(4-n)} - \beta_{\,3-n} \big)
\:\: , \qquad
\eeq
where $\beta_{n}$ are the N$^n$LO coefficients of the beta-function 
of QCD, and therefore read
\bea
\label{eq:C4+D4}
C_{\rm g}^{\,(4)} &\!=\!&
%
   \colourcolour{\caf} \* \left( 
    {177664 \over 81} 
   - {4288 \over 3}\,\*\zeta_2 
   + {704 \over 3}\,\*\zeta_3 
   + {1728 \over 5}\,\*\zeta_2^2 
   \right)
 + \colourcolour{\nf\*\cf\*\cas} \* \left( 
   - {880 \over 3} 
   + 256\,\*\zeta_3 \right)
\nn \\[0.5mm] && \mbox{\hspn}
 + \colourcolour{\nf\*\cat}\* \left( 
   - {41504 \over 81} 
   + {640 \over 3}\,\*\zeta_2 
   - {896 \over 3}\,\*\zeta_3 \right)
 + {1216 \over 81}\,\*\colourcolour{\nfs\*\cas}
\; ,
\\[1mm]
D_{\rm g}^{\,(4)} &\!=\!&
%
   \colourcolour{\caf} \* \left( 
   - {1984 \over 27} 
   + 16\,\*\zeta_2 
   + 1072\,\*\zeta_3 
   + {88 \over 3}\,\*\zeta_2^2 
   - 160\,\*\zeta_3\*\zeta_2 
   - 320\,\*\zeta_5 \right)
\nn \\[0.5mm] && \mbox{\hspn}
 + \colourcolour{\nf\*\cat}\* \left( 
     {2048 \over 27} 
   - 16\,\*\zeta_2 
   - 160\,\*\zeta_3 
   - {16 \over 3}\,\*\zeta_2^2 \right)
 - 8\,\* \colourcolour{\nf\*\cf\*\cas}
 - {64 \over 27}\,\* \colourcolour{\nfs\*\cas} 
\; ,
\eea
Their numerical values in QCD, rounded to seven significant digits, 
are
\bea
\label{C4gNum}
  C_{\rm g}^{\,(4)} &\!=\!& 
        85814.12 \,-\, 13880.52\,\*\nf \,+\, 135.1111\,\*\nfs
\\
  D_{\rm g}^{\,(4)} &\!=\!&
        54482.81 \,-\, 4341.134\,\*\nf \,-\, 21.13333\,\*\nfs 
\; .
\label{D4gNum}
\eea
The corresponding result for the coefficient of the plus 
distribution $1/(1-x)_+$ in eq.~(\ref{eq:xto1}) reads
\beq
\label{A4gNum}
  A_{\rm g}^{\,(4)} \:=\: 
        40880.33 \,-\, 11714.25\,\*\nf \,+\, 440.0488\,\*\nfs 
  \,+\, 7.362775\,\*\nft 
\; .
\eeq
 
The quark and gluon quantities $B_{\rm k}^{\,(n)}$, which are sometimes
referred to as the virtual anomalous dimensions, are related via 
collinear anomalous dimensions $f_{\,\rm k}$ which exhibit a 
generalized Casimir scaling \cite{Moch:2018wjh}, for details see 
refs.~\cite{Das:2019btv,Das:2020adl}. 
Using the later results for the complete $1/\ep$ pole terms of the 
quark and gluon form factor \cite{Agarwal:2021zft,Lee:2021uqq}, 
also the $\nfz$ part of $B_{\rm g}^{\,(4)}$ can be related to 
$B_{\rm q}^{\,(4)}$ which is exactly known in the large-$\nc$ limit 
\cite{Moch:2017uml}.
This results in
\bea
\label{eq:B4g}
B_{\rm g}^{\,(4)} &\!=\!&
%
   \colourcolour{\caf} \* \left(
            {50387 \over 486}
          + {2098 \over 27}\,\*\z2
          + {48088 \over 27}\,\*\z3
          + {1793 \over 27}\,\*\zz(2,2)
          - {3902 \over 9}\,\*\z3\,\*\z2
          - {14617 \over 9}\,\*\z5
\right. \nn \\[0.5mm] && \left. \mbox{\hspp}
          - {76516 \over 945}\,\*\zz(2,3)
          + {682 \over 3}\,\*\zz(3,2)
          + {336 \over 5}\,\*\z3\,\*\zz(2,2)
          + 80\,\*\z5\,\*\z2
          + 700\,\*\z7
          - \frac{1}{24}\,\* b_{4,\, \dfFA}^{\,\rm q} 
   \right)
\nn \\[0.5mm] && \mbox{\hspn}
 + \colourcolour{\dfAAna}\* \left(
          - {800 \over 9}
          + {1184 \over 3}\,\*\z2
          - {784 \over 3}\,\*\z3
          - {1016 \over 15}\,\*\zz(2,2)
          - 272\,\*\z3\,\*\z2
          + {760 \over 3}\,\*\z5
\right. \nn \\[0.5mm] && \left. \mbox{\hspp}
          + {5984 \over 315}\,\*\zz(2,3)
          + b_{4,\, \dfFA}^{\,\rm q}
   \right)
  \:\: + \; \nf \mbox{ contributions }.
\eea
The terms suppressed here for brevity can be found in eq.~(10) of
ref.~\cite{Das:2020adl}. The present result for the remaining 
numerical coefficient, which drops out in the large-$\nc$ limit, 
is $\, b_{4,\, \dfFA}^{\,\rm q} \:=\: 998.02 \,\pm\, 0.02\, $
\cite{MVV-prp}.
The corresponding uncertainty of the $\nfo$ contribution is now
practically negligible; the coefficients of  $\nfs$ and $\nft$ were 
exact already in ref.~\cite{Das:2020adl}. The numerical value in 
QCD reads
\beq
\label{B4gNum}
  B_{\rm g}^{\,(4)} \;=\;
        68587.64 \pm 0.3 
  \,-\, 18143.98\,\*\nf 
  \,+\, 423.8113\,\*\nfs
  \,+\, 0.9067215\,\*\nft
\:\: .
\eeq
Recall that here, as before, the expansion is written in powers of 
$\ars = \als/(4\pi)$.

The small-$x$ behaviour of the flavour-singlet splitting functions is
dominated by the BFKL single-logarithmic enhancement of the $1/x$ 
contributions with leading-log (LL) terms of the form 
$\as(n)\, x^{-1} \ln^{\,n-1} x\,$ which vanish for $n=2,\,3$ and 5 
\cite{Jaroszewicz:1982gr,CataniFM90}. Thus the present four-loop order 
is the first that exhibits leading BFKL logarithms. 
So far the next-to-leading logarithmic (NLL) correction has been 
calculated \cite{Fadin:1998py,Ciafaloni:1998gs} and transformed to the
\MSb\ scheme \cite{Ciafaloni:2005cg,Ciafaloni:2006yk} only for 
$P_{\rm gg}$.

At three loops, the NLL BFKL contributions to $P_{\rm gq}$ and
$P_{\rm gg}$ are related by Casimir scaling in the large-$\nc$ limit; 
in QCD the breaking of this relation is numerically small, 
see eq.~(4.29) of ref.~\cite{Vogt:2004mw}. 
In view of the large uncertainties due to the unknown NNLL 
$x^{\,-1} \ln x$ terms, see below, assuming Casimir scaling for the 
$x^{\,-1} \ln^{\,2} x$ term of $P_{\rm gq}^{\,(3)}$ should not 
generate any relevant bias.

\begin{figure}[t]
\vspace{-2mm}
\centerline{\epsfig{file=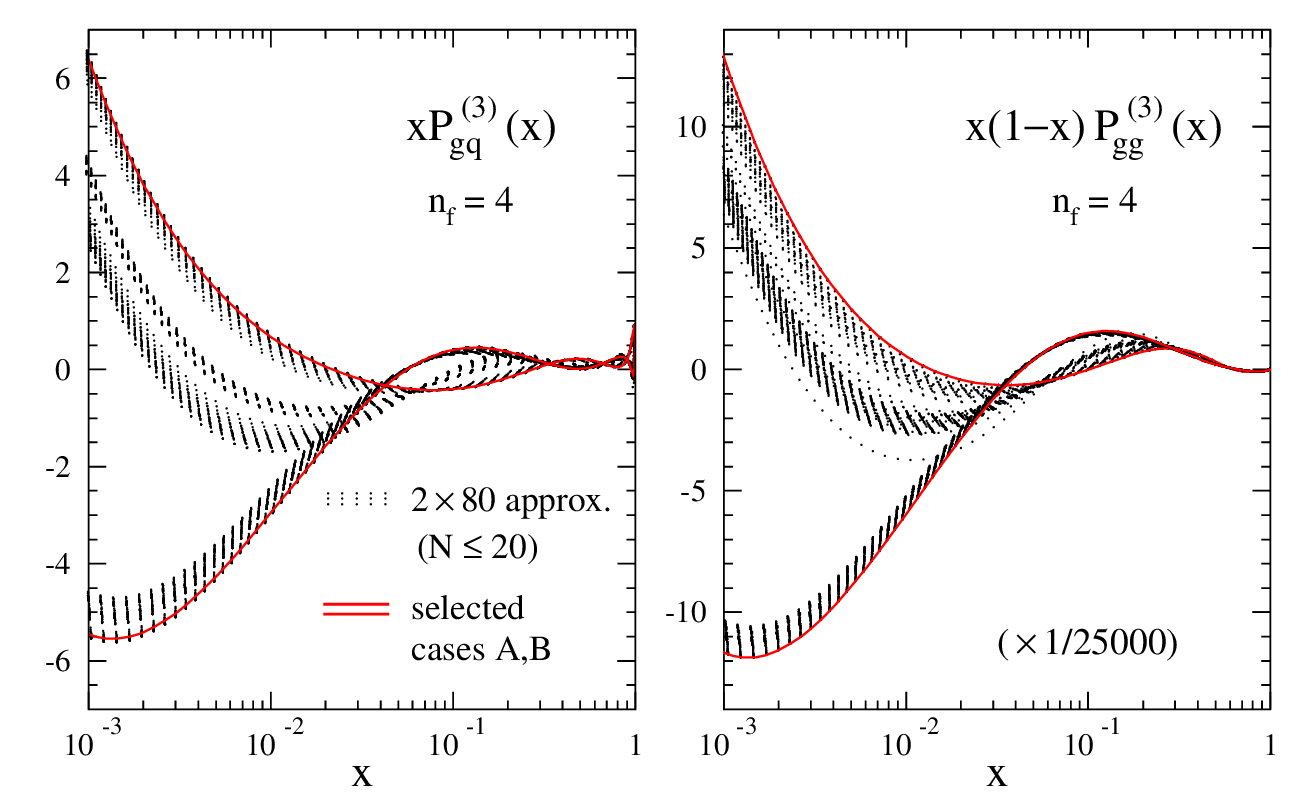,width=16.0cm,angle=0}}
\vspace{-3mm}
\caption{\label{fig:pgx3ab} \small
Two sets of 80 trial functions, one for a large and one for a small
value of the unknown coefficient of $x^{\,-1} \ln x$, for the
four-loop (N$^3$LO) contribution to the gluon-quark (left) and 
gluon-gluon (right) splitting functions at $\nf = 4$.
The two cases selected for eqs.~(\ref{eq:Pgq3A3-nf4}) and
(\ref{eq:Pgg3A3-nf4}) are shown by the solid (red) lines.}
\vspace{-2mm}
\end{figure}

We are now ready to present the $x$-space approximations constructed
using the lowest five even moments and the endpoint information.
We have first determined a range for the critical coefficient of
$x^{\,-1} \ln x$ that does not lead to excessive variations at 
large $x$ and aims to be sufficiently conservative at small $x$.
Then we built about 80 approximations for the boundaries of this range, 
using $\xm/x$, a~set of one-parameter polynomials, $\ln x$ or a 
combination with $\ln^{\,2} x$, and two of the three large-$x$ terms 
$\Lnt(a)$ (for $P_{\rm gg}$ with a prefactor $\xm$) with $a=1,2,3$.
Finally two representatives were selected for each of the lower-row 
splitting functions at the physically relevant numbers
$\nf = 3,\,4,\,5$ of light flavours. This process is illustrated in 
fig.~\ref{fig:pgx3ab} for $\nf = 4$.
 
The approximations are decomposed into 
\bea
\label{eq:Pgi-def}
 P_{\rm gi,\,A/B}^{\,(3)}(\nf,x) \,=\,
 p_{{\rm gi},0}^{\,[\nf]}(x) + p_{\rm gi,\,A/B}^{\,[\nf]}(x)
\,
\quad \mbox{ for } \quad {\rm i \,=\, q,g}
\; .
\eea
$p_{{\rm gi},0}^{\,(\nf)}(x)$ collects the known endpoint contributions, 
again rounded to seven significant figures,
\bea
\label{eq:Pgq30-nf}
{\lefteqn{ \hspn
 p_{{\rm gq},0}^{\,[\nf]}(x) \; = \; 
 -3692.719\,\*L_0^3/x
 -(47516.44+442.8369\,\*\nf)\,\*L_0^2/x
}}
\nn \\ && \mbox{}
 +(13.44307-0.5486968\,\*\nf)\,\*L_1^5
 +(375.3983-34.49474\,\*\nf+0.8779150\,\*\nfs)\,\*L_1^4
\, ,
\quad 
\\
\label{eq:Pgg30-nf}
{\lefteqn{ \hspn
 p_{{\rm gg},0}^{\,[\nf]}(x) \; = \; 
 - 8308.617\,\*L_0^3/x
 - (106912.0+996.3830\,\*\nf)\,\*L_0^2/x 
 + P_{\rm gg,\,x\ra 1\,}^{\,(3)}(x) 
}}
\eea
with $P_{\rm gg,\,x\ra 1\,}^{\,(3)}(x)$ from eq.~(\ref{eq:xto1})
with the coefficients in eqs.~(\ref{C4gNum})$\,$-$\,$(\ref{A4gNum})
and (\ref{B4gNum}), where we disregard the very small uncertainty 
of the latter.
In view of the small number of moments, we have not included the 
$x^{\,0}$ double logarithms of ref.~\cite{Davies:2022ofz}.

The selected approximate contributions to eq.~(\ref{eq:Pgi-def}) read
\bea
\label{eq:Pgq3A3-nf3}
 p_{\rm gq,\,A}^{\,[3]}(x) &\!=\!&
           - 166073\*L_0/x
           - 161562\* (1-x)/x
           + 36469
           + 72317\*L_0
           - 3977.3\*L_1^2
           +  484.4\*L_1^3
\, ,
\nn\\
 p_{\rm gq,\,B}^{\,[3]}(x) &\!=\!&
           - 263763\*L_0/x
           - 546482\* (1-x)/x
           - 39464
           - 401000\*L_0
           + 13270\*L_1^2
           + 3289\*L_1^3
\, ,
\nn\\
\\
\label{eq:Pgq3A3-nf4}
 p_{\rm gq,\,A}^{\,[4]}(x) &\!=\!& 
           - 167578\*L_0/x
           - 158805\* (1-x)/x
           + 35098
           + 87258\*L_0
           - 4834.1\*L_1^2
           +  176.6\*L_1^3
\, ,
\nn\\
 p_{\rm gq,\,B}^{\,[4]}(x) &\!=\!& 
           - 266154\*L_0/x
           - 547215\* (1-x)/x
           - 41523
           - 390350\*L_0
           + 12571\*L_1^2
           + 3007\*L_1^3
\, ,
\nn\\
\\
\label{eq:Pgq3A3-nf5}
 p_{\rm gq,\,A}^{\,[5]}(x) &\!=\!& 
           - 169084\*L_0/x
           - 154336\* (1-x)/x
           + 33889  
           + 103440\*L_0
           - 5745.8 \*L_1^2
           -  128.6\*L_1^3
\, ,
\nn\\
 p_{\rm gq,\,B}^{\,[5]}(x) &\!=\!& 
           - 268545\*L_0/x
           - 546236\* (1-x)/x
           - 43421
           - 378460\*L_0
           + 11816\*L_1^2
           + 2727.3\*L_1^3
\, ,
\nn \\
\eea
and
\bea
\label{eq:Pgg3A3-nf3}
 p_{\rm gg,\,A}^{\,[3]}(x) &\!=\!& 
           - 373663.9\*L_0/x
           - 345063\* (1-x)/x 
           + 86650\* (1+x^2)\*(1-x)
\nn \\[-0.5mm] && \mbox{}
           + 158160\*L_0
           - 15741\* (1-x)\*L_1^2
           - 9417\* (1-x)\*L_1^3  
\, ,
\nn\\[0.5mm]
 p_{\rm gg,\,B}^{\,[3]}(x) &\!=\!&
           - 593466.2\*L_0/x
           - 1265632\* (1-x)/x 
           - 656644\* (1+x^2)\*(1-x)
\nn \\[-0.5mm] && \mbox{}
           - 1352233\*L_0
           + 203298\* (1-x)\*L_1^2
           + 39112\* (1-x)\*L_1^3  
\, ,
\\[0.5mm]
\label{eq:Pgg3A3-nf4}
 p_{\rm gg,\,A}^{\,[4]}(x) &\!=\!& 
           - 377051.6\*L_0/x
           - 342625\* (1-x)/x 
           + 100372\* (1+x^2)\*(1-x)
\nn \\[-0.5mm] && \mbox{}
           + 189167\*L_0
           - 29762\* (1-x)\*L_1^2
           - 12102\* (1-x)\*L_1^3  
\, ,
\nn\\[0.5mm]
 p_{\rm gg,\,B}^{\,[4]}(x) &\!=\!& 
           - 598846.6\*L_0/x
           - 1271540\* (1-x)/x 
           - 649661\* (1+x^2)\*(1-x)
\nn \\[-0.5mm] && \mbox{}
           - 1334919\*L_0
           + 191263\* (1-x)\*L_1^2
           + 36867\* (1-x)\*L_1^3  
\, ,
\\[0.5mm]
\label{eq:Pgg3A3-nf5}
 p_{\rm gg,\,A}^{\,[5]}(x) &\!=\!& 
           - 380439.3\*L_0/x
           - 337540\* (1-x)/x 
           + 119366\* (1+x^2)\*(1-x)
\nn \\[-0.5mm] && \mbox{}
           + 223769\*L_0
           - 45129\* (1-x)\*L_1^2  
           - 15046\* (1-x)\*L_1^3
\, ,
\nn\\[0.5mm]
 p_{\rm gg,\,B}^{\,[5]}(x) &\!=\!& 
           - 604227.1\*L_0/x
           - 1274800\* (1-x)/x 
           - 637406\* (1+x^2)\*(1-x)
\nn \\[-0.5mm] && \mbox{}
           - 1314010\*L_0
           + 177882\* (1-x)\*L_1^2
           + 34362\* (1-x)\*L_1^3  
\, .
\eea

\begin{figure}[p]
\vspace{-2mm}
\centerline{\epsfig{file=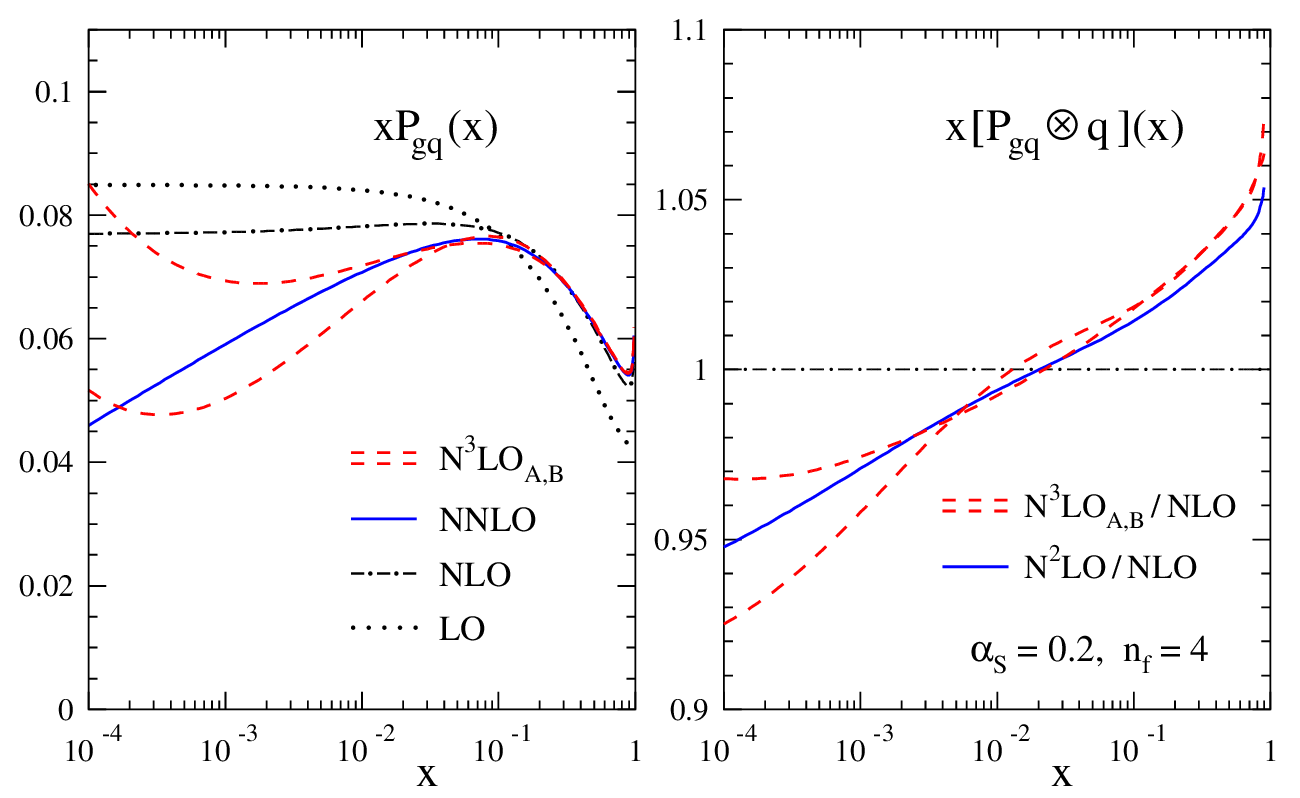,width=16.0cm,angle=0}}
\vspace{-3mm}
\caption{\label{fig:pgqn3lo} \small
Left: successive approximations to the \MSb\ quark-gluon splitting
function $P_{\rm gq}$, including the terms up to $\ar(n+1)$ for the 
N$^n$LO results, for $\nf = 4$ light flavours at $\als = 0.2$.
Right: the convolutions of the N$^2$LO and N$^3$LO approximations 
with the quark PDF in eq.~(\ref{qsgInp}), normalized to the NLO result.
}
\vspace{-2mm}
\end{figure}
\begin{figure}[p]
\vspace{-2mm}
\centerline{\epsfig{file=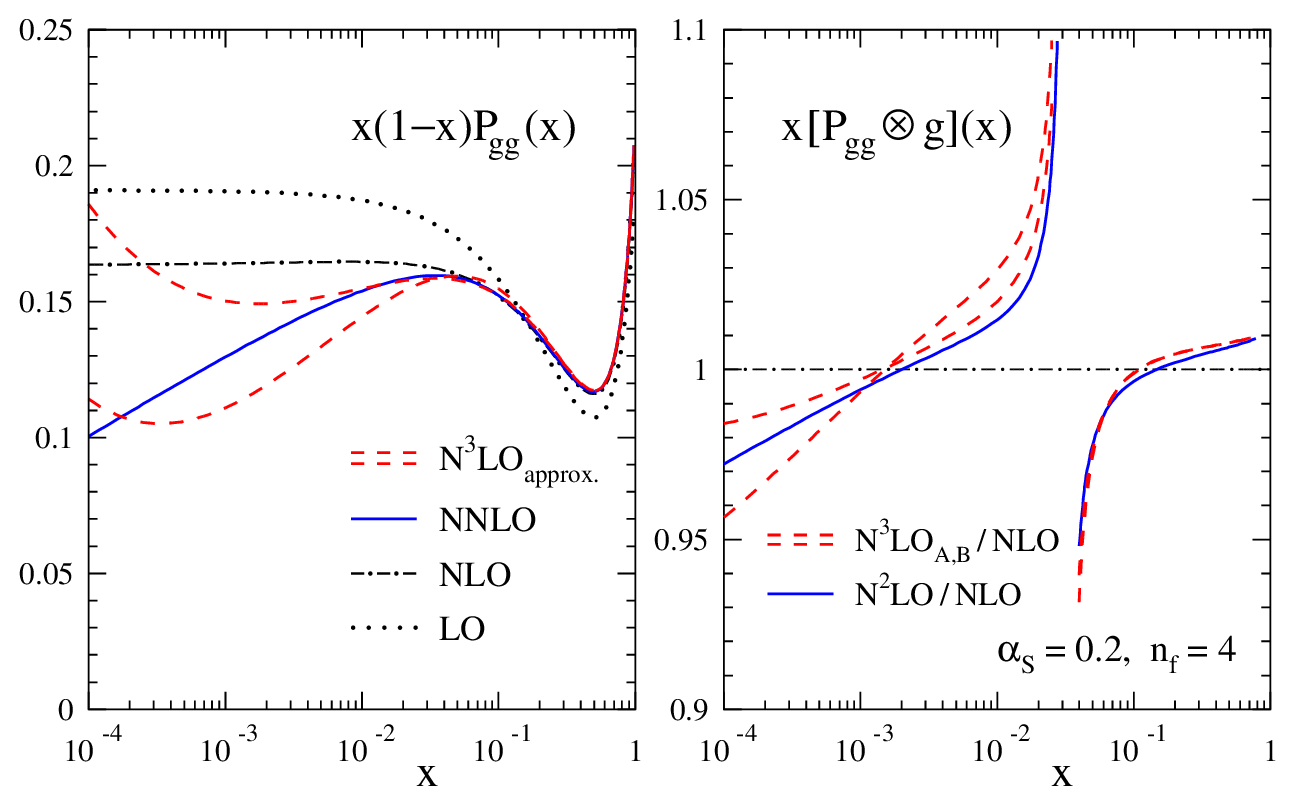,width=16.0cm,angle=0}}
\vspace{-3mm}
\caption{\label{fig:pggn3lo} \small
As fig.~\ref{fig:pgqn3lo}, but for the splitting function 
$P_{\rm gg}$, on the right convoluted with the gluon PDF in 
eq.~(\ref{qsgInp}).
}
\vspace{-2mm}
\end{figure}

In the left parts of figs.~\ref{fig:pgqn3lo} and \ref{fig:pggn3lo} 
we show the resulting perturbative expansions of the splitting 
functions $P_{\rm gq}$ and $P_{\rm gg}$ in eq.~(\ref{sgEvol}) at 
a standard reference point $\mu_0^{}$ with $\als(\mu_{0}^{\,2}) = 0.2$ 
and $\nf=4$.
The splitting functions themselves are well constrained only at
$x > 10^{\,-2}$. Of course, these functions enter the evolution
equations only via convolutions with quark or gluon PDFs.
The effect of these convolutions is illustrated in the respective
right panels for a schematic but sufficiently realistic
model input, already used in ref.~\cite{Vogt:2004mw},
\bea
\label{qsgInp}
  xq_{\sf s}^{}(x,\mu_{0}^{\,2}) &\!=\!&
  0.6\: x^{\, -0.3} (1-x)^{3.5}\, \left(1 + 5.0\: x^{\, 0.8\,} \right)
\: , \nn \\[-0.5mm]
  x\:\!g (x,\mu_{0}^{\,2})\:    &\!=\!&
  1.6\: x^{\, -0.3} (1-x)^{4.5}\, \left(1 - 0.6\: x^{\, 0.3\,} \right)
\; .
\eea
The convolutions lead to a considerable wider range of small 
uncertainties which extends to $x \gsim 10^{\,-3}$.
It is not yet possible to say whether the N$^3$LO corrections
are positive or negative below $x \simeq 10^{\,-3}$ at this 
reference point, yet one can safely expect that they will amount to
no more than a few percent even at $x = 10^{\,-4}$. 

\vspace{2mm}
To summarize, we have computed the even-$N$ moments up to $N=10$ of all
4-loop (N$^3$LO) flavour-singlet splitting functions for the evolution
of unpolarized parton distributions (PDFs) of hadrons. 
These analytic computations have been performed, using the 
{\sc Forcer} program~\cite{Ruijl:2017cxj} in {\sc Form}
\cite{Vermaseren:2000nd,Kuipers:2012rf,Ruijl:2017dtg}, 
via the conceptually straightforward but computationally very 
challenging route of structure functions in inclusive deep-inelastic 
scattering \cite{Larin:1993vu,Larin:1996wd}.

The hardest (meta-)$\,$diagrams for the gluon-quark and gluon-gluon
cases required more than \mbox{$2\cdot 10^{\,8}$} CPU seconds on 
state-of-the-art multi-core workstations at $N=10$, their storage 
demand at the peak of the intermediate expression swell reached well
above 20 TB. This excludes extending the present computations to $N=12$ 
by any incremental improvements of our present means.

Fortunately there is another approach functions, via off-shell operator 
matrix elements, that allows to continue the determination of the 
4-loop splitting functions to higher values of $N$. This approach,
however, is conceptionally much more involved for flavour-singlet
quantities in non-Abelian gauge theories such as QCD.
The present results can employed in the process of validating such 
calculations, as already done for the pure-singlet and gluon-quark case
in refs.~\cite{Falcioni:2023luc,Falcioni:2023vqq}. We hope that those 
computations will be extended to the quark-gluon and gluon-gluon 
cases fairly soon.

In the meantime, approximate N$^3$LO analyses of the proton's
PDFs and of benchmark processes can be performed using the results of 
refs.~\cite{Falcioni:2023luc,Falcioni:2023vqq} and the present paper.
For this purpose, we have constructed approximate results for the
N$^3$LO splitting functions $P_{\rm gq}^{\,(3)}(x)$ and
$P_{\rm gg}^{\,(3)}(x)$ that, due to the smoothening effect of the
Mellin convolutions, should have an acceptable accuracy for PDFs
at momentum fractions $x \gsim 10^{\,-3}$. 
An important ingredient for the gluon-gluon case is that, as 
demonstrated above, all contributions to $P_{\rm gg}^{\,(3)}(x)$ 
that do not vanish for $x \ra 1$ are now practically (if not yet 
analytically, for one coefficient) known.

\vspace{3mm}
\noindent
{\sc Form} and {\sc Fortran} file with our results for 
$\gamma_{\,\rm ik}^{\,(3)}(N)$ at $N\!=\!8$ and $N\!=\!10$, 
the large-$x$ coefficients in $P_{\rm gg, x\ra 1}^{\,(3)}(x)$, 
and the numerical approximations for $P_{\rm ik}^{\,(3)}(x)$ 
have been deposited at the preprint server {\tt https://arXiv.org} 
with the sources of this letter. $\phantom{10^10}$
%
%
\subsection*{Acknowledgements}
\vspace*{-1mm}

This work has been supported by the Deutsche Forschungsgemeinschaft through 
Research Unit FOR 2926, {\it Next Generation pQCD for Hadron Structure: 
Preparing for the EIC}, project number 40824754, and DFG grant MO~1801/4-2; 
the ERC Advanced Grant 101095857 {\it Conformal-EIC};
the Japan Society for the Promotion of Science (JSPS) under the 
KAKENHI Grant Numbers 19K03831, 21K03583 and 22K03604;
and the Consolidated Grant ST/T000988/1 of the Science and Technology 
Facilities Council (STFC), United Kingdom.

{\small
\addtolength{\baselineskip}{-1.5mm}

\providecommand{\href}[2]{#2}\begingroup\raggedright\endgroup

}

\end{document}